\documentclass[twocolumn,reprint,superscriptaddress,showpacs,floatfix,amsmath,amssymb,aps,pra]{revtex4-1}

\usepackage{graphicx}
\usepackage{dcolumn}
\usepackage{bm}
\usepackage{braket}
\usepackage{longtable}
\usepackage{placeins} 
\usepackage{hyperref}
\usepackage{color}
\usepackage[colorinlistoftodos]{todonotes}

\def\pulsenum{10}
\def\thtos{10.4}
\def\thtop{32.2}
\def\exptos{7.7(1)}
\def\exptop{29.9(1)}

\newcommand{\tj}[6]{ \begin{Bmatrix}
  #1 & #2 & #3 \\
  #4 & #5 & #6 
\end{Bmatrix}}

\begin{document}

\title{Anisotropic dependence of tune-out wavelength near Dy 741-nm transition}

\author{Wil Kao}
\affiliation{Department of Applied Physics, Stanford University, Stanford, California 94305, USA}
\affiliation{E.~L.~Ginzton Laboratory, Stanford University, Stanford, California 94305, USA}
\author{Yijun Tang}
\affiliation{E.~L.~Ginzton Laboratory, Stanford University, Stanford, California 94305, USA}
\affiliation{Department of Physics, Stanford University, Stanford, California 94305, USA}
\author{Nathaniel Q. Burdick}
\affiliation{Department of Applied Physics, Stanford University, Stanford, California 94305, USA}
\affiliation{E.~L.~Ginzton Laboratory, Stanford University, Stanford, California 94305, USA}
\author{Benjamin L. Lev}
\affiliation{Department of Applied Physics, Stanford University, Stanford, California 94305, USA}
\affiliation{E.~L.~Ginzton Laboratory, Stanford University, Stanford, California 94305, USA}
\affiliation{Department of Physics, Stanford University, Stanford, California 94305, USA}

\begin{abstract}
We report the first measurement of a tune-out wavelength for ground-state bosonic Dy and linearly polarized light.  The tune-out wavelength measured is near the narrow-line 741-nm transition in $^{162}$Dy, and is the wavelength at which the total Stark shift of the ground state vanishes.  We find that it strongly depends on the relative angle between the optical field and quantization axis due to Dy's large tensor polarizability.  This anisotropy provides a wide, 22-GHz tunability of the tune-out frequency for linearly polarized light, in contrast to Rb and Cs whose near-infrared tune-out wavelengths do not exhibit large anisotropy.  The measurements of the total light shift are performed by measuring the contrast of multipulse Kapitza-Dirac diffraction.  The calculated wavelengths are within a few GHz of the measured values using known Dy electronic transition data.  The lack of hyperfine structure in bosonic Dy implies that the tune-out wavelengths for the other bosonic Dy isotopes should be related to this $^{162}$Dy measurement by the known isotope shifts.
\end{abstract}

\pacs{
}
 \maketitle

\section{Introduction}
The recent trapping and laser cooling of magnetic dipolar atomic elements such as chromium~\cite{Bradley2000}, erbium~\cite{McClelland2006}, dysprosium~\cite{Lu:2010fo,Youn2010}, thulium~\cite{Sukachev2010}, and holmium~\cite{Miao2014}, with the first three having been cooled to quantum degeneracy~\cite{Griesmaier2005,Lu2011bec,Lu2012,Aikawa2012,Aikawa:2014if,Tang2015,Naylor:2015bs}, has opened new avenues of ultracold atomic physics exploration.
Specifically, the long-range and anisotropic character of the magnetic dipole-dipole interaction provides a platform to investigate the role dipolar physics can play in quantum simulation.   Examples of the latter include proposals to realize topologically non-trivial systems~\cite{Zeng2015,Yao2015}, and recent progress includes the study of the extended Bose-Hubbard model using erbium~\cite{Baier:2016ga} and the observation of the arrested implosion of a dipolar dysprosium BEC due to the balance between the mean-field potential and quantum fluctuations~\cite{Kadau:2016cb,Wachtler:2016gb,Baillie:2016cc,FerrierBarbut2016,Chomaz:2016ua,Schmitt:2016wg}.

Neutral atoms  experience a force in an inhomogeneous light field.  The resulting trapping force arises from the interaction between the light field and the induced atomic dipole moment and leads to the so-called Stark shift in the atomic energy level~\cite{Grimm:2000ed}.
The total Stark shift can vanish at certain wavelengths due to cancellation between multiple atomic transitions. 
Such ``tune-out" wavelengths for various atomic species have been predicted theoretically~\cite{LeBlanc2007,Arora2011,Cheng2013,Dammalapati2016} and measured experimentally~\cite{Herold2012,Henson2015,Clark2015,Schmidt2016}. 
This knowledge is particularly useful for engineering species-specific trapping potentials. 
For example, in mixed-species experiments, one can create an optical lattice potential for one species but not the other \cite{Catani2009,McKay2013,Topcu2013,Vaidya2015}, and the interaction between the trapped species and the background species allows for the implementation of novel cooling schemes to realize new quantum phases~\citep{Griessner2006}. 
This species-specific lattice can also be used to tune the interspecies effective mass ratio of the trapped atoms~\cite{LeBlanc2007}, allowing for the flexible exploration of collective dynamics~\cite{Kramer2002,Pu2003}.
Lastly, the coexistence of trapped fermions and a background bosonic gas can potentially introduce phonon-like excitations to an optical lattice~\cite{Ott2004,LeBlanc2007,Bruderer2010}.

Aside from engineering trapping potentials for neutral atoms, knowledge of tune-out wavelengths is also useful in the context of the optical control of Feshbach resonances (OFR), a promising technique to achieve time-varying and/or spatially modulated interatomic interactions~\cite{Bauer2009,Yan2013,Fu2013,Clark2015}.
One can operate the OFR laser at a far-off-resonance regime to reduce heating and loss rate and, if possible, at the tune-out wavelength of the targeted atomic species so that the parasitic dipole force of the OFR beam is eliminated, as recently demonstrated by Ref.~\cite{Clark2015}.
The longer lifetime permits studies of the non-equilibrium dynamics of quantum gases with long observation times when the tune-out wavelength is far-detuned from electronic transitions~\cite{Clark2015}.

The precise determination of atomic transition strengths cannot rest on \textit{ab initio} quantum-mechanical calculations alone due to the electronic complexity of lanthanides like dysprosium, the atom considered here.  
The complicated electronic structure of these open-$f$-shell lanthanide elements presents a significant challenge for such analyses (see Ref.~\cite{Dzuba2011} and the recent study Ref.~\cite{Li:2016vy}) and experimental investigations are crucial for understanding their electronic structure.
For example, in Ref.~\cite{Lepers2014}, a semi-empirical approach that utilizes both theoretical calculations and experimental data leads to the prediction of nine unobserved odd-parity energy levels in the erbium atomic spectrum.  More generally, improved knowledge of atomic polarizabilities, which are informed by measurements of tune-out wavelengths~\cite{Schmidt2016}, can guide choices of, e.g., optical dipole trapping wavelengths and laser wavelengths for implementing Raman transitions for realizing synthetic gauge fields~\cite{Lin:2009dr,Lin:2011hn,Burdick:2016jt}.  It is in this spirit that we present the measurement of the tune-out wavelength for the bosonic $^{\text{162}}$Dy near the narrow 741-nm transition. 

We proceed by describing the experimental system in Sec.~\ref{expmeth} before introducing the calculation of Stark shifts in Sec.~\ref{sec:theory} and results of our measurements in Sec.~\ref{results}.

\begin{figure}[t!]
\centering
\includegraphics[width=0.8\columnwidth]{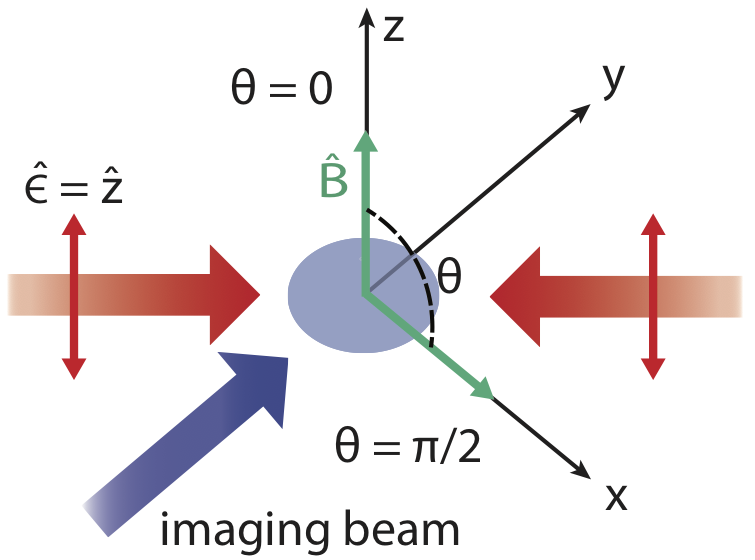}
\caption{Schematic of lattice beam (red arrow; with polarization $\hat{\epsilon}$) geometry used to measure lattice depth by KD diffraction of the BEC (blue sphere).  Green arrows indicate direction of the applied magnetic field.}
\label{fig:setup}
\end{figure}

\section{Experimental Methods}\label{expmeth}
Using linearly polarized light, we probe the total (scalar plus tensor) light shift in the vicinity of the 741-nm transition by Kapitza-Dirac (KD) lattice diffraction~\cite{Kapitza1933}, which is a standard tool for optical lattice characterization~\cite{Jo2012,Cheiney2013}. See  Fig.~\ref{fig:setup}. The vector light shift vanishes for linearly polarized light.  See Sec.~\ref{sec:theory}. The lattice depth $U_0$ is measured in units of recoil energy $E_r = (\hbar k_r)^2/2m$, where $m$ is the mass of the atom, $k_r=2\pi/\lambda$ is the grating wavevector, and $\lambda$ is the wavelength of the laser.

We perform KD diffraction such that $\omega_r t \gtrsim 1$, where the recoil angular frequency is $\omega_r = E_r/\hbar$ and the grating pulse length $t$ is sufficiently long to induce a coherent oscillation between the momentum states~\cite{Gadway2009}. After turning off the light grating diabatically, the atoms project into momentum states $\ket{2n\hbar k_r}$, where $n = 0, \pm 1, \pm 2, ...$ denotes the $n^{\text{th}}$ order of the diffracted matter wave. In the weak-lattice limit, $U_0 \ll 4E_r$, only orders with $|n| \le 1$ effectively participate in the coherent evolution, and the tune-out wavelength can be identified by decreasing population $P_1$ in the first-order diffraction peaks as $U_0$ approaches zero.

\begin{figure}[t!]
\centering
\includegraphics[width=\columnwidth]{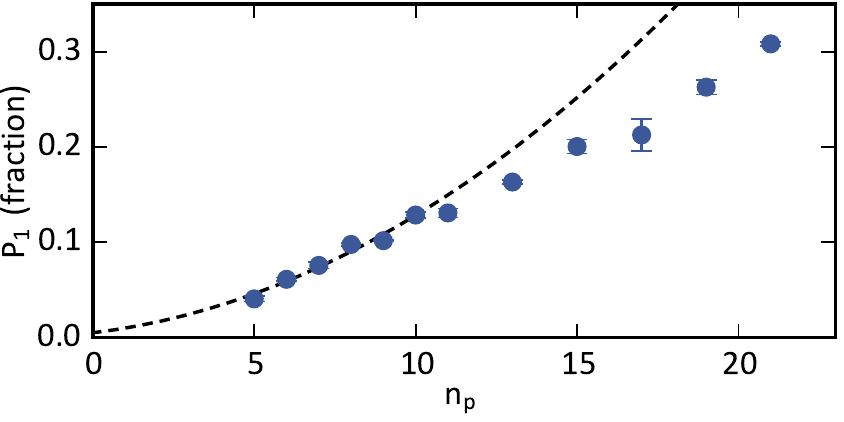}
\caption{Fractional first-order diffracted population at a constant laser power and wavelength.
Each point is an average of three measurements shown with 1$\sigma$ standard error.
For optimal SNR, we use the maximum pulse number up to which the quadratic enhancement holds (i.e., $n_p = 10$).
 The first six data points are fit to Eq.~\eqref{eq:depth}.  The dashed line is the resulting fit, which yields a measured lattice depth of $U_0 = 0.20E_R$.}
\label{fig:np}
\end{figure}

We employ a multipulse diffraction scheme to enhance the signal by constructive interference, as demonstrated in Ref.~\cite{Herold2012}.
We pulse an optical lattice along one spatial dimension near the 741-nm transition with a detuning $\Delta=(\omega_L-\omega_0)/(2\pi)$, where $\omega_L$ is the laser frequency and $\omega_0$ is the resonant frequency of the 741-nm transition.
In the weak lattice limit, the oscillation period between $\ket{0}$ and $\ket{\pm 2\hbar k_r}$ approaches $\tau = h/(4E_r)=111~\mu$s for $^{162}$Dy.
As in  Ref.~\cite{Herold2012}, we use a square-wave sequence with $n_p$ pulses since there is an $n_p^2$ enhancement in $P_1$ in the weakly diffracting limit ($n_p U_0 \ll 4E_r$) given by 
\begin{equation}\label{eq:depth}
P_1 = \frac{n_p^2 |U_0|^2}{32E_r^2},
\end{equation}
where the sign of $U_0$ can be determined by the sign of the detuning $\Delta$. The period is 111~$\mu$s with a 50\% duty cycle.
Finally, we probe a range of $\Delta$ in search of the tune-out wavelength.

The quadratic scaling of $P_1$ with respect to $n_p$ in Eq.~\eqref{eq:depth} no longer holds after the diffracted peaks of atoms have sufficiently moved in space to lose coherence---and thus the ability to constructively interfere---with the main condensate. This coherence time limit from overlap loss sets the maximum pulse number where the quadratic scaling is valid. As shown in Fig.~\ref{fig:np}, the $P_1$ enhancement efficiency starts to drop below quadratic at $n_p = \pulsenum$, after which the atoms in $\ket{\pm 2\hbar k_r}$ have traveled $62$\% of the Thomas-Fermi radius. 
Therefore, for accurate light shift measurements, we employ a ten-pulse sequence to amplify $P_1$.
Because the coherence time only depends on $k_{r}$, this choice of $n_p$ is appropriate for all $\Delta$ used in our measurements.

The tune-out wavelength is measured in terms of the detuning, denoted $\Delta_0$, from the 741-nm transition frequency. 
We use a wavelength meter \footnote{HighFinesse WS/6 200} to monitor the frequency of the lattice beam derived from a Ti:Sapphire laser and to calibrate the wavelength meter against our frequency-stabilized 741-nm diode cooling laser with a $<$10~kHz/h drift rate~\cite{Lu2011bec} via a beat note setup. 
While simultaneously monitoring the frequency of the lattice beam and the cooling laser on the wavelength meter, we combine beams of both lasers onto a high-speed photodetector \footnote{Electro-Optics Technology ET-4000}.
We then compare the beat signal frequency, measured up to $\pm 8$~GHz, to the frequency difference given by the wavelength meter. 
Applying this calibration beyond the range of $\pm 8$~GHz yields an uncertainty of 10~MHz in the frequency measurement.

We prepare Bose-Einstein condensates (BECs) of $^{162}$Dy using methods described in Ref.~\cite{Tang2015}.
The resulting trap frequencies are $\left[f_x, f_y, f_z\right] = \left[62(2), 32(4), 113(2)\right]$~Hz, where gravity is along $\hat{z}$. 
We initiate KD diffraction with a nearly pure BEC of $5 \times 10^4$ atoms in the maximally stretched ground state $\ket{J=8,m_J=-8}$.
As illustrated in Fig.~\ref{fig:setup}, the 1D lattice is formed along $\hat{x}+\hat{y}$ by retroreflecting a $0.20(2)$-W collimated beam with a diameter of $950~\mu$m. 
The light field polarization is kept linear along $\hat{z}$, purified by a polarizing beam splitter.
Therefore,  any anisotropy in $\Delta_0$ should be attributed to the tensor light shift since the vector light shift is identially zero for linearly polarized light (see second term in Eq.~\eqref{eq:shift} below).
To probe the anisotropy in $\Delta_0$, we perform the measurement at two different field orientations $\hat{z}$ and $\hat{x} + \hat{y}$, both at a field magnitude of 1.580(5)~G, to realize $\theta = 0$ and $\theta = \pi/2$ in Eq.~\eqref{eq:shift}.
We note that this field is away from any Feshbach resonances~\cite{Baumann2014}.

\begin{figure}[t!]
\centering
\includegraphics[width=\columnwidth]{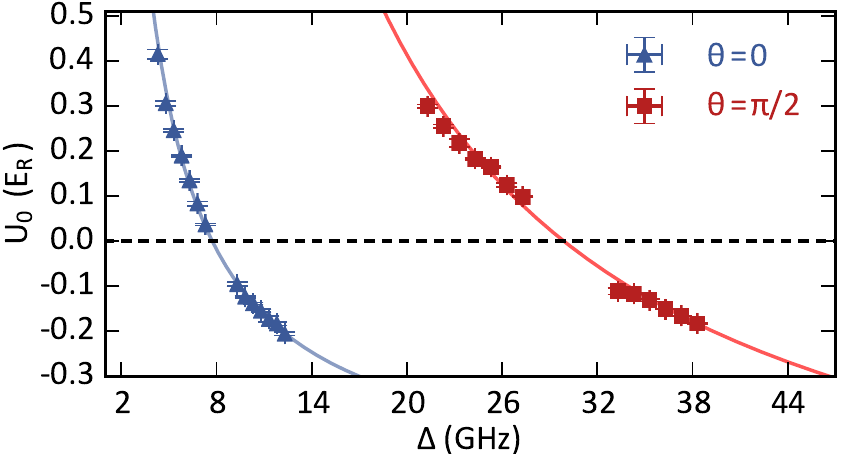}
\caption{Experimentally measured lattice depth versus detuning from the 741-nm transition for $\theta = 0$ (triangles) and $\theta = \pi/2$ (squares).
Each point is an average of three measurements, and the lines are fits to the expected functional form of the Stark shift from Eq.~\eqref{eq:form}.  Error bars are 1$\sigma$ standard error.}
\label{fig:exp}
\end{figure}

\begin{figure*}[t!]
\centering
\includegraphics[width=\linewidth]{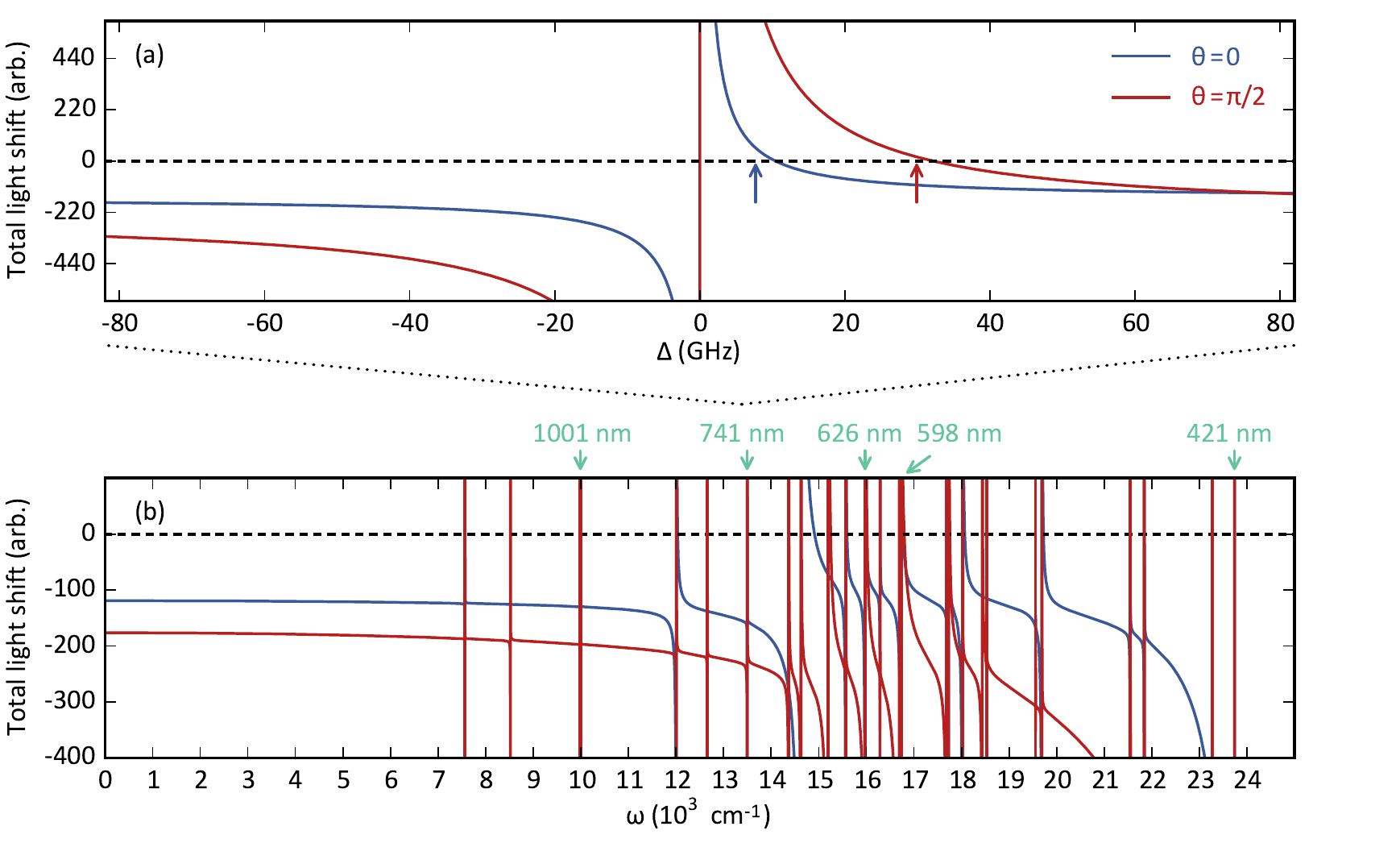}
\caption{(a) Calculated total light shift (arbitrary unit) versus detuning from the 741-nm line for both $\theta = 0$ and $\theta = \pi/2$.
The arrows indicate the positions of the measured $\Delta_0$ points. 
The 160-GHz window corresponds to 5~cm$^{-1}$.
(b) Calculated total light shift (arbitrary unit) for both polarizations between laser frequencies (wavelengths) DC and 24,000~cm$^{-1}$ (417~nm).
The cycling transitions ($J \rightarrow J' = J+1$) are indicated in turquoise \cite{Youn2010,Lu2011}. The large anisotropy at far-off resonant frequencies is likely a result of incomplete knowledge of atomic transition data; see main text for details.
}
\label{fig:theory}
\end{figure*}

\section{Stark shift calculation}\label{sec:theory}
For a single transition from the ground state $\ket{F,m_{F}}$ to an excited state $\ket{F^{'},m_{F^{'}}}$, the light shift of the ground state with an applied optical field of angular frequency $\omega$ is given by
\begin{widetext}
\begin{equation}\label{eq:shift}
\begin{aligned}
\Delta U_{(F,m_F)} &= -\alpha^{(0)}\lvert\vec{E}_0^{(+)}\rvert^2 -\alpha^{(1)} \left[i \vec{E}_0^{(-)} \times \vec{E}_0^{(+)}\right]_z \frac{m_F}{F} - \alpha^{(2)} \lvert\vec{E}_0^{(+)}\rvert^2 \left(\frac{3\cos^2\theta - 1}{2}\right) \left[\frac{3m_F^2-F(F+1)}{F(2F-1)}\right],\\
\alpha^{(0)} &= \frac{2\omega_{FF'}\lvert\bra{F}\lvert\vec{d}\rvert\ket{F'}\rvert^2}{3\hbar(2F+1)(\omega_{FF'}^2-\omega^2)},\\
\alpha^{(1)} &= (-1)^{F+F'+1}\sqrt{\frac{6F}{(F+1)(2F+1)}}\tj{1}{1}{1}{F}{F}{F'}\frac{\omega_{FF'}\lvert\bra{F}\lvert\vec{d}\rvert\ket{F'}\rvert^2}{\hbar(\omega_{FF'}^2-\omega^2)},\\
\alpha^{(2)} &= (-1)^{F+F'} \sqrt{\frac{40F(2F-1)}{3(F+1)(2F+1)(2F+3)}}\tj{1}{1}{2}{F}{F}{F'}  \frac{\omega_{FF'}\lvert\bra{F}\lvert\vec{d}\rvert\ket{F'}\rvert^2}{\hbar(\omega_{FF'}^2-\omega^2)},
\end{aligned}
\end{equation}
\end{widetext}
where $ \vec{E}_0^{(+)}$ and $\vec{E}_0^{(-)}$ are the rotating and counter-rotating part of the optical field; $\theta$ is the angle between its polarization and the quantization axis set by the magnetic field; $\omega_{FF'}$ is the transition angular frequency; $\alpha^{(0)}$, $\alpha^{(1)}$, and $\alpha^{(2)}$ are the scalar, vector and tensor polarizabilities; and $\lvert\bra{F}\lvert\vec{d}\rvert\ket{F'}\rvert^2$ is the reduced dipole matrix element~\cite{Deutsch1998,Geremia2006,Mitroy2010,Steck_QOnotes}. As in  Ref.~\cite{Dzuba2011}, we add a constant offset of $\alpha'=94.27$ atomic units (a.u.) to the calculated scalar polarizability to match the latest experimentally measured scalar polarizability at DC for $^{162}$Dy~\cite{Lei2015}. A similar correction for the tensor polarizability may be necessary but is yet unmeasured.
From Eq.~\eqref{eq:shift}, we note that while the scalar part preserves spherical symmetry, the vector light shift is dependent on the cross product of the rotating and counter-rotating electric fields and therefore vanishes if the polarization is linear.
The tensor term also breaks spherical symmetry with a $\theta$-dependence such that it is maximized at $\theta = 0$ and minimized at $\theta = \pi/2$.

The light shift of alkali atoms at large detuning is due solely to the scalar shift and is spherically symmetric.  For these atoms, when the laser detuning is large compared to the hyperfine splitting of the excited state, the hyperfine levels become effectively degenerate and the optical field interacts directly with the fine-structure transition.  That is, the hyperfine transition $F \to F'$ is replaced by the fine-structure transition $J \to J'$, $\omega_{FF'}$ is replaced by $\omega_{JJ'}$, and the polarizabilities can be directly expressed by the fine-structure dipole matrix elements. 
For alkali atoms in the ground state $\ket{J = 1/2, L = 0}$, the tensor polarizability $\alpha^{(2)}$ vanishes at large detuning, and the vector polarizability $\alpha^{(1)}$ is canceled by opposite contributions from the D$_1$ and D$_2$ lines~\cite{Deutsch1998,Geremia2006,Schmidt2016}. 

This spherical symmetry does not hold for Dy from the ground state $\ket{J = 8, L = 6}$ due to its large orbital angular momentum (bosonic Dy has zero nuclear spin).
For example, the tensor polarizability $\alpha^{(2)}_{741}$ for the 741-nm $J=8 \to J'=9$ transition does not vanish like the alkali $J=1/2 \to J'$ transitions.
On the contrary, it is on the same order of magnitude as the scalar polarizability $\alpha^{(0)}_{741}$ for all detunings ($\alpha^{(2)}_{741}\approx -0.7\alpha^{(0)}_{741}$).

\section{Results}\label{results}
We measure $P_1$ and determine $U_0$ using Eq.~\eqref{eq:depth}. 
Fig.~\ref{fig:exp} shows the measured $U_0$ with respect to the detuning $\Delta$ from the 741-nm resonance for both $\theta = 0$ and $\theta = \pi/2$.
We find that $P_1$ decreases as the laser is detuned further on the blue side of the 741-nm transition. 
Eventually, $P_1$ drops below the noise floor in our absorption image. 
As we further increase $\Delta$, we observe a revival in $P_1$, indicating a sign change in the total light shift.
We observe a clear polarization-dependence in $\Delta_0$, which differs by more than 20~GHz for the two polarizations.

We perform a least-squares fit to the data to quantitatively determine $\Delta_0$, using
\begin{equation}\label{eq:form}
U_0 = A + \frac{B}{\Delta}
\end{equation}
as the fit model, where $A$ and $B$ are free parameters.
Due to the large number of  dysprosium lines, it is intractable at present for us to use the exact analytical form of the total light shift as the fit function.
However, since the detuning from all lines other than the 741-nm transition is large, we simply treat the contribution from these other lines as a constant background.
The fitted curves cross zero at $\Delta_{0}(\theta=0) = \exptos$~GHz and $\Delta_{0}(\theta=\pi/2) = \exptop$~GHz, resulting in a 22.2-GHz  polarization-dependent anisotropy in $\Delta_0$.  The ratio of this anisotropy to the detuning from the 741-nm resonance is nearly five orders of magnitude larger than ratio of Rb's tune-out wavelength anisotropy ($\sim$44~MHz) to the detuning ($\sim$10~nm) from the D$_1$ and D$_2$ lines~\cite{Schmidt2016}.

We also compare the measured $\Delta_0$ values to the calculated Stark shift in Fig.~\ref{fig:theory}. 
We compute the total Stark shift using the formalism presented in Section~\ref{sec:theory}.
To account for all the known transitions of dysprosium, we sum over Eq.~\eqref{eq:shift} for the 26 lines documented in Ref.~\cite{Dzuba2011}, replacing the listed theoretical matrix elements with experimentally measured values when possible (e.g., the 741-nm transition~\cite{Lu2011}).
The calculated values---$\Delta_{0}^{\text{th}}(\theta=0) = \thtos$~GHz and $\Delta_{0}^{\text{th}}(\theta=\pi/2) = \thtop$~GHz---differ from the measured values by 26\% and 7\%, respectively.

Furthermore, while Ref.~\cite{Lu2011} reports an experimentally measured linewidth of 1.78(2)~kHz for the 741-nm excited state, we extract a linewidth by setting the reduced matrix element for the 741-nm transition $\mu_{741}$ as the free parameter and minimize the error function 
\begin{equation}
E(\mu_{741}) = \sum_{\theta=0,\pi/2}\left[\Delta_{0}^{\text{th}}(\theta)-\Delta_{0}(\theta)\right]_{\mu_{741}}^2, 
\end{equation}
which is the discrepancy between the calculated tune-out wavelengths $\Delta_0^{\text{th}}$, evaluated at $\mu_{741}$, and the experimental values $\Delta_0$ at $\theta = 0$ and $\theta = \pi/2$.
Despite the lack of the comprehensive measurements of all matrix elements of dysprosium~\cite{Dzuba2011}, we obtain 1.62(1)~kHz, which differs from the experimental value reported in Ref.~\cite{Lu2011} by 10\%.

Figure~\ref{fig:theory} suggests that the total light shift at $\theta = 0$ and $\theta = \pi/2$ should differ by around 50\% near 1064~nm (9,398~cm$^{-1}$), a common wavelength for optical dipole traps, and the one employed here.
To explore this, we measure the trap frequency along $\hat{z}$ for fields along $\hat{z}$ and $\hat{x}$, which correspond to $\theta = 0$ and $\theta = \pi/2$, respectively. 
Nevertheless, we find that the trap shape is isotropic to within 3\%, in contradiction to the prediction.
This isotropy has also been predicted in the optical trapping of erbium, another lanthanide atom, at 1064~nm using 1284 lines, 33 of which have been observed experimentally~\cite{Lepers2014}. See also the recent paper Ref.~\cite{Li:2016vy} for dysprosium calculations. In summary, while our calculation shows good agreement when sufficiently close to a resonance (less than tens-of-GHz-detuned), a more complete knowledge of the dysprosium atomic spectrum may be required to properly account for the far-off-resonance regime.

\section{Conclusions}
The measurement of the tune-out wavelengths for $^{\text{162}}$Dy near the narrow 741-nm transition has been presented, along with a calculation that reproduces these wavelengths within a few GHz.  As also predicted, we observe an anisotropy in the tune-out wavelength as a function of the relative angle between the optical field polarization and the quantization axis.  

While this work focuses on the tune-out wavelengths of $^{162}$Dy, those of the other bosonic isotopes of Dy are related to $^{162}$Dy's by the isotope shifts---listed in Ref.~\cite{Lu2011}---since nuclear effects play little role within the resolution of our measurement and bosonic Dy is $I=0$.  However, the wavelengths of the fermionic isotopes cannot be accurately determined with this data since those isotopes do possess hyperfine structure on the GHz scale~\footnote{Analogous measurements using Kapitza-Dirac diffraction for (near) degenerate fermions is much more challenging due to the intrinsically larger momentum spread.}. 

\begin{acknowledgements}
We thank V.~Vaidya for valuable discussions and acknowledge support from the AFOSR and NSF.
Y.T.~acknowledges partial support from the Stanford Graduate Fellowship.
\end{acknowledgements}

%
\end{document}